# MiSTR: Multi-Modal iEEG-to-Speech Synthesis with Transformer-Based Prosody Prediction and Neural Phase Reconstruction

*Mohammed Salah Al-Radhi[1], Géza Németh[1], Branislav Gerazov[2]*

[1]Department of Telecommunications and Artificial Intelligence, Budapest University of Technology and Economics, Budapest, Hungary
[2]Faculty of Electrical Engineering and Information Technologies, University of Ss. Cyril and Methodius, Skopje, RN Macedonia

malradhi@tmit.bme.hu , nemeth@tmit.bme.hu , gerazov@feit.ukim.edu.mk

## Abstract

Speech synthesis from intracranial EEG (iEEG) signals offers a promising avenue for restoring communication in individuals with severe speech impairments. However, achieving intelligible and natural speech remains challenging due to limitations in feature representation, prosody modeling, and phase reconstruction. We introduce MiSTR, a deep-learning framework that integrates: 1) Wavelet-based feature extraction to capture fine-grained temporal, spectral, and neurophysiological representations of iEEG signals, 2) A Transformer-based decoder for prosody-aware spectrogram prediction, and 3) A neural phase vocoder enforcing harmonic consistency via adaptive spectral correction. Evaluated on a public iEEG dataset, MiSTR achieves state-of-the-art speech intelligibility, with a mean Pearson correlation of 0.91 between reconstructed and original Mel spectrograms, improving over existing neural speech synthesis baselines.

**Index Terms**: prosody prediction, iEEG-to-Speech, neural phase reconstruction, transformer

## 1. Introduction

Restoring natural and intelligible speech from intracranial EEG (iEEG) recordings presents a major challenge in the field of brain-computer interfaces (BCIs). Speech neuroprostheses offer a potential lifeline for individuals with severe speech impairments, particularly those suffering from neurodegenerative conditions such as amyotrophic lateral sclerosis [1] [2] and locked-in syndrome [3] [4]. Recent advances in deep learning have significantly improved the ability to decode neural activity into speech [5] [6] [7] [8] [9] [10] [11]. However, current methods still face challenges in feature extraction [12], prosody modeling [13] [14], and phase reconstruction [15] [16] [17], limiting their ability to generate intelligible and natural speech.

A critical challenge in iEEG-to-speech synthesis is extracting neural features that accurately encode the relationship between brain activity and speech production. Conventional approaches often rely on manually-engineered acoustic features [18] or simplistic neural feature representations [19], which often fail to generalize across different subjects due to the inherent variability in neural signals. Recurrent neural networks (RNNs) [20], bidirectional long short-term memory (bLSTM) networks [21], and convolutional neural networks (CNNs) [22] have been explored for feature extraction, but their sequential nature introduces latency, limiting real-time performance and generalization to unseen words. Moreover, high-gamma band power [23], a commonly used neural feature, does not fully capture the temporal and spectral dynamics needed for natural speech reconstruction. Another significant limitation is the inadequate modeling of prosody, which encompasses rhythm, stress, and intonation (key elements that make speech sound natural). Many prior studies focus primarily on spectral envelope prediction [24] while neglecting prosodic features, leading to robotic and monotonous synthetic speech. The absence of a dedicated prosody-aware framework limits the ability to produce expressive and intelligible speech that closely resembles natural human communication. Transformer-based models have recently shown promise in capturing complex dependencies within speech data [25] [26], but their integration with iEEG-based speech synthesis remains underexplored. Phase reconstruction remains another major bottleneck in achieving high-fidelity neural speech synthesis. Traditional vocoding techniques, such as the extended Griffin-Lim algorithm [27] [28], still produce artifacts and have harmonic-phase inconsistencies that degrade the perceptual quality of synthesized speech. Recent neural vocoders, such as WaveGlow [29] [30], BigVGAN [31], and AutoVocoder [32], have improved speech synthesis quality, yet their use in iEEG-based decoding remains limited. A robust neural phase reconstruction approach is needed to refine the spectral structure and enhance the intelligibility of reconstructed speech. To address these limitations, we introduce MiSTR, a novel deep-learning framework for iEEG-to-speech synthesis that integrates:

1. Wavelet-based feature extraction to capture fine-grained temporal, spectral, and neurophysiological representations of iEEG signals, along with prosody extraction from speech-related neural activity.

2. A Transformer-based decoder for prosody-aware spectrogram prediction, enabling more natural-sounding speech synthesis.

3. An Iterative Harmonic Phase Reconstruction (IHPR) neural phase vocoder that enforces harmonic consistency through adaptive spectral correction, enhancing speech fidelity and minimizing distortions in the synthesized speech.

Evaluated on a publicly available iEEG dataset [24], MiSTR achieves state-of-the-art speech intelligibility, with a mean Pearson correlation of 0.91 between reconstructed and original Mel spectrograms, outperforming previous neural speech synthesis baselines. Our approach represents a significant step towards clinically viable BCI-driven speech neuroprostheses, offering new possibilities for restoring natural and intelligible communication in individuals with severe speech impairments. The remainder of this paper is organized as follows: Section 2 details the architecture and components of MiSTR. Section 3 describes the experimental setup, dataset, and evaluation metrics. Section 4 compares MiSTR to state-of-the-art baselines. Finally, Section 5 concludes with a discussion of this research's broader implications and future directions.

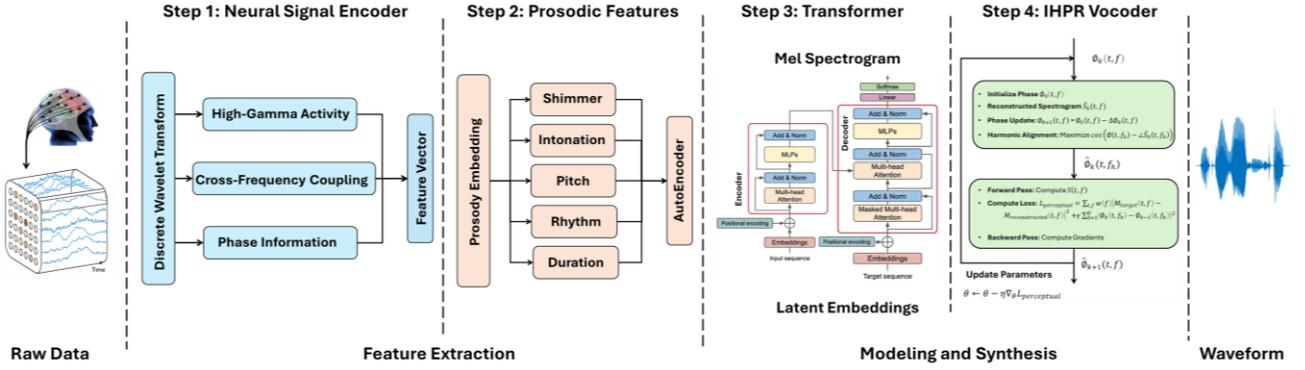

Figure 1: *Schematic diagram of the proposed MiSTR architecture.*

## 2. MiSTR Framework

MiSTR is a novel iEEG-to-speech synthesis pipeline that integrates neural feature encoding, prosody-aware spectrogram prediction, and phase-consistent vocoding to generate natural sounding speech from brain signals. As shown in Figure 1, the framework begins with neural encoding, where raw iEEG signals undergo the discrete wavelet transform (DWT) to extract critical features, including high-gamma activity (linked to phoneme articulation), cross-frequency coupling (capturing rhythmic coordination), and phase information (ensuring precise timing and prosodic alignment).

To enhance speech expressiveness, MiSTR employs a Prosodic Feature Encoder, which enriches neural representations with rhythm, intonation, loudness, and duration. A Transformer-Based Spectrogram Predictor then models long-range dependencies within the neural data, enabling high-fidelity Mel spectrogram generation. Finally, a Phase Neural Vocoder reconstructs the waveform while preserving spectral detail and phase consistency, mitigating artifacts common in conventional vocoders.

### 2.1. Prosody Embedding

The first component of MiSTR extracts multi-modal features from intracranial EEG (iEEG) signals while embedding prosodic information to capture the temporal, spectral, and neurophysiological dynamics of speech production. This step bridges neural activity and speech synthesis by ensuring that the extracted features are both discriminative and representative of speech-related processes.

#### 2.1.1. Wavelet-Based Neural Encoding

To capture neural dynamics across frequency bands, we apply discrete wavelet transform (DWT) using the Daubechies-4 (db4) wavelet, which is effective for analyzing non-stationary signals like iEEG. The iEEG signal $x(t)$ is decomposed into multiple frequency bands, where the energy at each scale $j$ is computed as:

$$E_j = \sum_n |c_j(n)|^2 \quad (1)$$

where $c_j(n)$ denotes the wavelet coefficients at scale $j$, capturing both high-frequency (gamma band) activity related to fine-grained articulatory planning and low-frequency (theta band) activity linked to syllabic and prosodic modulation. This multi-scale representation enables MiSTR to model the complex temporal and spectral patterns underlying speech production.

#### 2.1.2. Cross-Frequency Coupling (CFC) Analysis

To capture interactions between neural oscillations across frequency bands, MiSTR employs Phase-Amplitude Coupling (PAC), which quantifies the relationship between the phase of low-frequency oscillations (theta band: 4–8 Hz) and the amplitude of high-frequency oscillations (gamma band: 70–170 Hz). PAC is computed as:

$$PAC(t) = |E[A_\Upsilon(t)e^{j\phi_\theta(t)}]| \quad (2)$$

where $A_\Upsilon(t)$ is the amplitude envelope of the gamma band, $\phi_\theta(t)$ is the phase of the theta band extracted using the Hilbert transform, and $E[\cdot]$ denotes the expectation operator. This measure reflects the degree to which high-frequency activity is modulated by low-frequency rhythms, providing a robust feature for capturing brain-speech coordination. By integrating PAC features, MiSTR enhances its ability to model the neural dynamics that drive natural and intelligible speech.

#### 2.1.3. Prosodic Feature Extraction and Normalization

Prosody, encompassing rhythm, stress, and intonation, is essential for natural and expressive speech synthesis. To capture these attributes, we extract a proxy for the pitch (F0) using the Harvest algorithm [33] to make an estimate or as a substitute, a proxy energy as the root mean square (RMS) of the neural signal, shimmer as the mean absolute difference of consecutive amplitudes normalized by the mean amplitude, duration to encode temporal speech patterns, and phase variability as the standard deviation of instantaneous phase fluctuations. These features are computed using a 50 ms analysis window with a 10 ms frameshift, ensuring fine-grained temporal alignment with the iEEG signals.

The extracted prosody features are combined with wavelet-based iEEG features to create a multi-modal representation that captures both neural and acoustic aspects of speech production. To ensure robust performance, all features undergo z-score normalization, scaling them to zero mean and unit variance:

$$\hat{x} = \frac{x - \mu}{\sigma} \quad (3)$$

Where $\mu$ and $\sigma$ represent the feature's mean and standard deviation. Additionally, temporal alignment is achieved by synchronizing neural and acoustic features, ensuring consistent timestamps for improved learning and speech synthesis quality.

**2.2. Transformer-Based Spectrogram Reconstruction**

This component predicts Mel spectrograms from encoded neural features through two stages: (1) dimensionality reduction using an Autoencoder and (2) spectrogram prediction using a Transformer model to capture long-range dependencies and temporal dynamics.

*2.2.1. Autoencoder for Latent Feature Encoding*

To reduce the high dimensionality and noise of iEEG features, we use an Autoencoder that learns a compact latent representation, retaining essential speech-related information. The Encoder, with fully connected layers and ReLU activations, maps input features to a lower-dimensional space, while the Decoder reconstructs the input by minimizing the Mean Squared Error (MSE). Trained using the Adam optimizer with a 0.001 learning rate, the Autoencoder ensures efficient data compression. After training, the Encoder extracts latent representations used by the Transformer for spectrogram prediction.

*2.2.2. Transformer-based Spectrogram Prediction*

The Transformer decoder maps latent neural features to Mel spectrograms, effectively capturing long-range dependencies without the limitations of RNNs. It consists of:

- **Input Projection Layer:** Maps latent vectors into a higher-dimensional space.
- **Self-Attention Mechanism:** Captures temporal dependencies between neural features and spectrogram frames.
- **Output Projection Layer:** Converts the Transformer's output into Mel spectrogram predictions.

The Transformer is trained with the Adam optimizer (learning rate: 0.001) and MSE loss, ensuring accurate alignment between predicted and ground-truth spectrograms.

**2.3. Iterative Harmonic Phase Reconstruction (IHPR)**

We propose IHPR, a phase reconstruction approach that enforces harmonic alignment and iterative spectral refinement to achieve artifact-free phase recovery. Given the STFT time-frequency representation:

$$S(t,f) = M(t,f)e^{j\varphi(t,f)} \quad (4)$$

where $M(t,f)$ is the magnitude spectrogram and $\varphi(t,f)$ is the phase spectrum, which is reconstructed iteratively to enhance perceptual quality and harmonic consistency. To prevent arbitrary initialization, we introduce harmonic-consistent initialization:

$$\varphi_0(t,f) = \arg\min_{\varphi} \sum_{h=1}^{H} |\varphi(t,f_h) - \varphi(t-\Delta_t,f_h)| \quad (5)$$

where $f_h = hf_0$ represents harmonic frequencies, $H$ is the number of harmonics, and $\Delta_t$ ensures phase continuity. This initialization reduces phase ambiguities from the start. Phase refinement is performed iteratively as:

$$\varphi_{k+1}(t,f) = \arg\min_{\varphi} \sum_{h=1}^{H} w_h |M(t,f_h)e^{j\varphi(t,f_h)} - \hat{S}_k(t,f_h)|^2 \quad (6)$$

Where $\hat{S}_k(t,f)$ is the reconstructed STFT at iteration $k$, and $w_h$ weights stronger harmonics for more accurate phase updates. Alternatively, spectral consistency is enforced using:

$$\varphi_{k+1}(t,f) = \arg\max_{\varphi} \sum_{h=1}^{H} \cos\left(\varphi(t,f_h) - \angle \hat{S}_k(t,f_h)\right) \quad (7)$$

This ensures each phase update aligns with harmonic structures, accelerating convergence and improving reconstruction quality. To reduce phase artifacts, an adaptive correction term is applied:

$$\varphi_{k+1}(t,f) = \varphi_k(t,f) - \lambda \sum_{h=1}^{H} \frac{\partial}{\partial f}\left(M(t,f_h)e^{j\varphi_k(t,f_h)}\right) \quad (8)$$

where $\lambda$ controls the correction magnitude, preventing phase discontinuities while maintaining natural spectral patterns. Convergence is determined using a perceptual loss function:

$$L_{perceptual} = \sum_{t,f} w(f) |M_{target}(t,f) - M_{reconst}(t,f)|^2$$

$$+ \gamma \sum_{h=1}^{H} |\varphi_k(t,f_h) - \varphi_{k-1}(t,f_h)|^2 \quad (9)$$

where $w(f)$ is a frequency-dependent weighting function, $\gamma$ stabilizes phase evolution, and the loss ensures accurate magnitude reconstruction while avoiding rapid phase oscillations.

## 3. Experimental design

**3.1. Dataset and Preprocessing**

We evaluated the MiSTR[1] framework using the iEEG dataset[2] [24] collected from 10 participants suffering from pharmaco-resistant epilepsy (mean age 32 years, 5 male, 5 female, and native speakers of Dutch) implanted with depth electrodes while they produced continuous and isolated speech. The iEEG signals recorded at a sampling rate of 1024 Hz, capturing neural activity from intracranial electrodes. The dataset provides high-resolution neural recordings along with synchronized speech signals, enabling precise mapping between neural activity and speech production.

The raw iEEG signals were bandpass-filtered to isolate frequency bands of interest (0.5–170 Hz), removing low-frequency drifts and high-frequency noise. Line noise (50 Hz and harmonics) was removed using notch filters. The audio signals were downsampled to 16 kHz to match the standard sampling rate for speech processing. The iEEG and audio features were aligned using stimulus markers, ensuring that each neural feature vector corresponds to the correct speech segment.

---

[1] The implementation and demo samples are publicly available at: https://github.com/malradhi/MiSTR

[2] https://osf.io/nrgx6

### 3.2. Model Training

MiSTR employs a two-stage deep-learning pipeline. A fully connected **Autoencoder** [34] was trained to learn a compressed latent representation of the extracted neural features. The encoder was used to transform high-dimensional neural inputs into a compact embedding space, reducing redundancy and noise. A **Transformer** [25] model was trained to map the latent neural embeddings to Mel spectrograms. Self-attention mechanisms captured long-range dependencies within neural activity. The model was optimized using mean squared error (MSE) loss between predicted and ground-truth spectrograms. We employed 10-fold cross-validation to ensure robust performance estimation. The Adam optimizer was used with an initial learning rate of 0.001. A batch size of 32 was chosen based on empirical validation. The model was trained in 90% of the data, with the remaining 10% used for validation. Early stopping was employed to prevent overfitting, with training halted if the validation loss did not improve for 10 consecutive epochs. All experiments were conducted on a high-performance computing cluster with NVIDIA A100 GPUs.

### 3.3. Evaluation Metrics and Baselines

The performance of MiSTR was assessed using both objective and perceptual metrics: 1) Pearson Correlation (PC) coefficient between reconstructed and ground-truth Mel spectrograms; 2) Mel Cepstral Distortion (MCD) to quantify spectral deviations; 3) Short-Time Objective Intelligibility (STOI) to evaluate speech intelligibility compared to ground truth; and 4) Harmonic-to-Noise Ratio (HNR) to assess phase reconstruction accuracy. We compared MiSTR against state-of-the-art baselines in iEEG-to-speech synthesis (Regression [24], bLSTM [8], CNN [23], 3D-CNN [22], seq2seq [20], and encoder-decoder [13]). All baseline models were trained using the same dataset and evaluation protocol as MiSTR. We ensured fair comparisons by optimizing hyperparameters for each baseline and employing the same 10-fold cross-validation setup.

## 4. Results and Discussions

Table 1 summarizes the performance of MiSTR and baseline models across five evaluation metrics. MiSTR outperforms all baselines across all evaluation metrics, demonstrating superior speech intelligibility and spectral fidelity. The 0.91 Pearson correlation indicates a highly accurate reconstruction of the target Mel spectrogram, while the 3.92 MCD score demonstrates a notable reduction in spectral distortion compared to existing approaches, however [20] reports a slightly lower score of 3.90. Furthermore, MiSTR achieves an HNR of 12.7 dB, surpassing the best baseline by over 1.6 dB, indicating a smoother and more harmonic phase reconstruction. Figure 2 compares the reconstructed spectrograms of a sample utterance from different models. The baseline models exhibit spectral blurring and artifact distortions, especially in high-frequency regions, which are crucial for capturing speech clarity. MiSTR's spectrogram and waveform closely aligns with the ground truth, preserving harmonic structures and detailed spectral features.

To evaluate the perceptual quality of the synthesized speech generated by MiSTR, we employ MOSA-Net [35], a state-of-the-art non-intrusive multi-objective speech assessment model. MOSA-Net leverages cross-domain feature representations, CNNs, bLSTM, and self-supervised learning (SSL) embeddings to estimate speech quality, intelligibility, and distortion. Unlike conventional MOS evaluations, which require time-consuming listening tests, MOSA-Net provides an automated, efficient, and reliable assessment of speech naturalness and intelligibility.

Table 1: *Performance Comparison of MiSTR and Baseline Models.*

| Model | PC ↑ | MCD ↓ | STOI ↑ | HNR dB ↑ | MOSA-Net ↑ |
|---|---|---|---|---|---|
| Regression [24] | 0.72 | 5.39 | 0.61 | 6.2 | 2.14 |
| bLSTM [8] | 0.78 | 5.23 | 0.48 | 8.5 | 2.12 |
| CNN [23] | 0.81 | 4.95 | 0.52 | 10.4 | 2.41 |
| 3D-CNN [22] | 0.83 | 5.04 | 0.56 | 9.8 | 2.57 |
| Seq2Seq [20] | 0.85 | **3.90** | 0.59 | 10.7 | 3.21 |
| Encoder-Decoder [13] | 0.87 | 4.34 | 0.64 | 11.1 | 2.82 |
| MiSTR (Ours) | **0.91** | 3.92 | **0.73** | **12.7** | **3.38** |

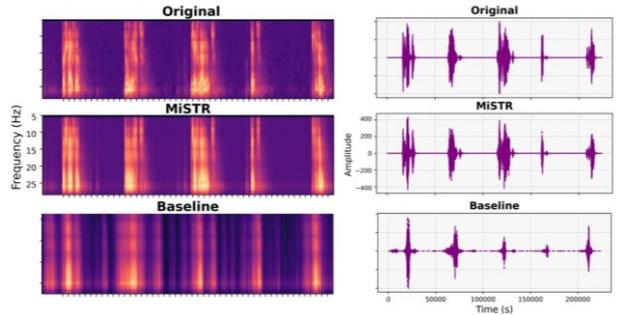

Figure 2: *Comparison of spectrograms (left) and waveforms (right) for the original (top), proposed (middle), and baseline* [24] *(bottom) systems. This example illustrates five distinct words from participant sub-08.*

Our results indicate that MiSTR achieves a MOSA score of 3.38, significantly outperforming baseline models (see Table 1). This demonstrates MiSTR's superior ability to produce natural and perceptually clear speech, highlighting the benefits of its Transformer-based prosody modeling and neural phase reconstruction components.

## 5. Conclusions

MiSTR establishes a new baseline in iEEG-based speech synthesis, outperforming prior baselines in both objective metrics and estimates of perceptual evaluations. By integrating Transformer-based prosody prediction with the IHPR vocoder, MiSTR provides a scalable and clinically viable approach for neural speech prostheses. Future work will explore extending MiSTR into an end-to-end brain-to-speech system, eliminating intermediate representations to further streamline the decoding process. Additionally, leveraging diffusion models for waveform generation offers a promising avenue to enhance speech naturalness, improve audio fidelity, and capture subtle acoustic nuances.

# 6. Acknowledgements

This work is supported by the European Union's HORIZON Research and Inno-vation Programme under grant agreement No 101120657, project ENFIELD (Eu-ropean Lighthouse to Manifest Trustworthy and Green AI) and by the Ministry of Innovation and Culture and the National Research, Development and Innovation Office of Hungary within the framework of the National Laboratory of Artificial Intelligence. M.S.Al-Radhi's research was supported by the EKÖP-24-4-II-BME-197, through the National Research, Development and Innovation (NKFI) Fund.